# Multitask Scanning Probe Microscopy

*Aditya Raghavan[1], Yu Liu[1], Ian Mercer[2], JP Maria[2], Sergei Kalinin[1]*
[1] University of Tennessee, Knoxville, TN, USA
[2]Pennsylvania State University, University Park, PA, USA

Scanning probe microscopy provides nanoscale access to structural, electrical, electromechanical, magnetic, and mechanical properties of materials. Its increasing use for wafer-scale characterization and combinatorial materials exploration creates a need to distribute measurements efficiently across large spatial domains. This is particularly important when available modalities differ in acquisition time and potential for tip and sample damage, making exhaustive multimodal mapping over spatial grids impractical. Here, we demonstrate multitask scanning probe microscopy, a live, closed-loop workflow in which a multitask Gaussian process learns spatial and cross-modal relationships and autonomously selects both the next measurement location and the next experimental protocol. The approach is implemented on an automated large-sample atomic force microscope and demonstrated on a composition-spread AlScN wafer using tapping-mode and Dual AC Resonance Tracking (DART) measurements. Paired initial measurements establish the relation between the tasks, after which noncoincident measurements are used to update both response landscapes. The resulting workflow extends active learning in scanning probe microscopy from spatial sampling to autonomous allocation of measurement modalities and provides a basis for combining rapid, weakly perturbative imaging with slower contact, electrical, electromechanical, magnetic, or spectroscopic measurements.

araghav4@utk.edu, sergei2@utk.edu

## I. Introduction

Scanning probe microscopy (SPM) has emerged as a key method for investigating materials and nanoscale phenomena.[1] In addition to the surface morphology accessible through contact and intermittent-contact imaging, scanning probe methods allow spatially resolved measurements of currents,[2] surface potential,[3] electromechanical response,[4] magnetic interactions,[5] mechanical properties,[6] and local spectroscopic responses to electrical and mechanical stimuli. The combination of high spatial resolution and direct access to local functional behavior has made scanning probe microscopy central to studies of semiconductors, ferroic and energy materials, polymers, biological systems, and nanoscale devices.[1] The same instrumental platform can often access several complementary properties of the material by changing the detection mode, excitation waveform, or by enabling a broad range of spectroscopic measurements.

Of particular interest is the application of scanning probe microscopy to complex systems such as large-area samples, including semiconductor wafers, composition-spread films, and combinatorial materials libraries.[7-11] These platforms enable exploration of thickness and process uniformity and spatial variation of functional properties, and discovery of composition-structure-property relationships.[10, 11] In the combinatorial-library geometry, a single sample can contain a continuous or discrete family of compositions, processing conditions, phases, and microstructures. The experimental challenge then shifts from imaging one selected region to distributing a finite measurement budget across a large spatial or compositional domain. The number of possible measurement locations can readily reach thousands, while the number of available measurement modalities and associated protocols introduces an additional dimension to the experiment.

The measurement protocols available in scanning probe microscopy differ considerably in their experimental cost and potential for tip and sample damage. Tapping-mode topography can provide rapid and comparatively weakly perturbative exploration of surface morphology.[1, 12] In contrast, contact-mode current imaging, local current-voltage spectroscopy, piezoresponse force microscopy, switching spectroscopy, and force-volume measurements are associated with considerably longer measurement times.[2, 4, 6, 13] Similarly, contact-mode measurements, especially with high probing biases, can accelerate tip wear and modify soft or weakly bound surfaces.[14] Functional measurements can therefore be both time-consuming and state-changing, particularly when repeated across a large wafer or combinatorial library.

The resulting experimental problem contains two coupled decisions. The microscope must determine where an additional measurement will be informative and which measurement protocol should be performed at that location. Conventional automated and autonomous scanning probe workflows have largely focused on the first of these decisions.[15, 16] In these approaches, a statistical model is updated continuously with incoming measurements and, directing the instrument toward the most informative next measurement location. Specific implementations have used Gaussian-process regression,[17, 18] deep kernel learning,[19] hypothesis learning,[20, 21] and image-based fixed-policy workflows[22, 23] to identify spatial regions, experimental parameters, or physical conditions for subsequent measurement.

In most implementations, however, the measurement protocol, i.e. the choice of which SPM mode to use, is selected before the experiment and remains fixed throughout the active-learning process. Multimodal measurements are consequently acquired according to a predefined sequence or repeated at every selected location, irrespective of whether all modalities provide sufficient additional information. Recent studies have begun to relax this constraint, demonstrating automated selection of the most informative observational channel in multimodal piezoresponse force microscopy[24] and multi-task co-orchestration of measurement location and modality on pre-acquired multimodal datasets.[25] Live, closed-loop selection and execution of both the measurement location and the instrument protocol on an operating microscope has, however, remained outstanding.

Here we introduce multitask scanning probe microscopy as a framework for live, closed-loop autonomous selection of both the measurement location and the measurement modality on an operating microscope. In this approach, individual scanning probe modalities are treated as related prediction problems, referred to as tasks in the statistical learning literature,[26, 27] each with its own spatial response landscape and measurement noise but sharing a common underlying material structure. A multitask Gaussian process learns the spatial structure within each task together with the correlations between tasks.[26, 27] A measurement performed using one modality updates the model predictions for that modality and, through the learned statistical correlation between modalities, simultaneously refines the predictions for the unmeasured modality across the entire wafer[25-27] (Fig. 1b). At each iteration, the controller selects a location-task pair, configures the microscope for the corresponding mode, executes the measurement, extracts the relevant response, and updates the joint model.

We implement this concept on an Oxford Instruments Asylum Research Jupiter atomic force microscope equipped for large-area wafer navigation and automated switching between tapping and contact-resonance modes. The method is demonstrated on a composition-spread AlScN wafer[28, 29] using roughness measured by tapping-mode topography[12] and by Dual AC Resonance Tracking (DART) imaging[30, 31] as two related tasks. This choice provides a controlled first implementation in which both tasks respond to the same underlying surface morphology while differing in probe-sample interaction, feedback conditions, tuning requirements, and measurement noise. More general formulations applicable to modality pairs with weaker or more complex correlations, such as topography paired with electrical or electromechanical measurements, are discussed as extensions of the framework.

## II. Multitask Gaussian Processes for Scanning Probe Microscopy

### II.A. Measurement Protocols as Statistical Tasks

The definition of a task is central to multitask scanning probe microscopy (Fig. 1). Here, a task corresponds to an independently selectable measurement protocol that consumes experimental time or requires a distinct configuration of the tip-sample interaction. Examples include tapping-mode imaging, contact current mapping, Kelvin probe force microscopy (KPFM), piezoresponse force microscopy (PFM), magnetic-force microscopy (MFM), local current-voltage spectroscopy, and switching spectroscopy PFM.[2-6, 12, 13]

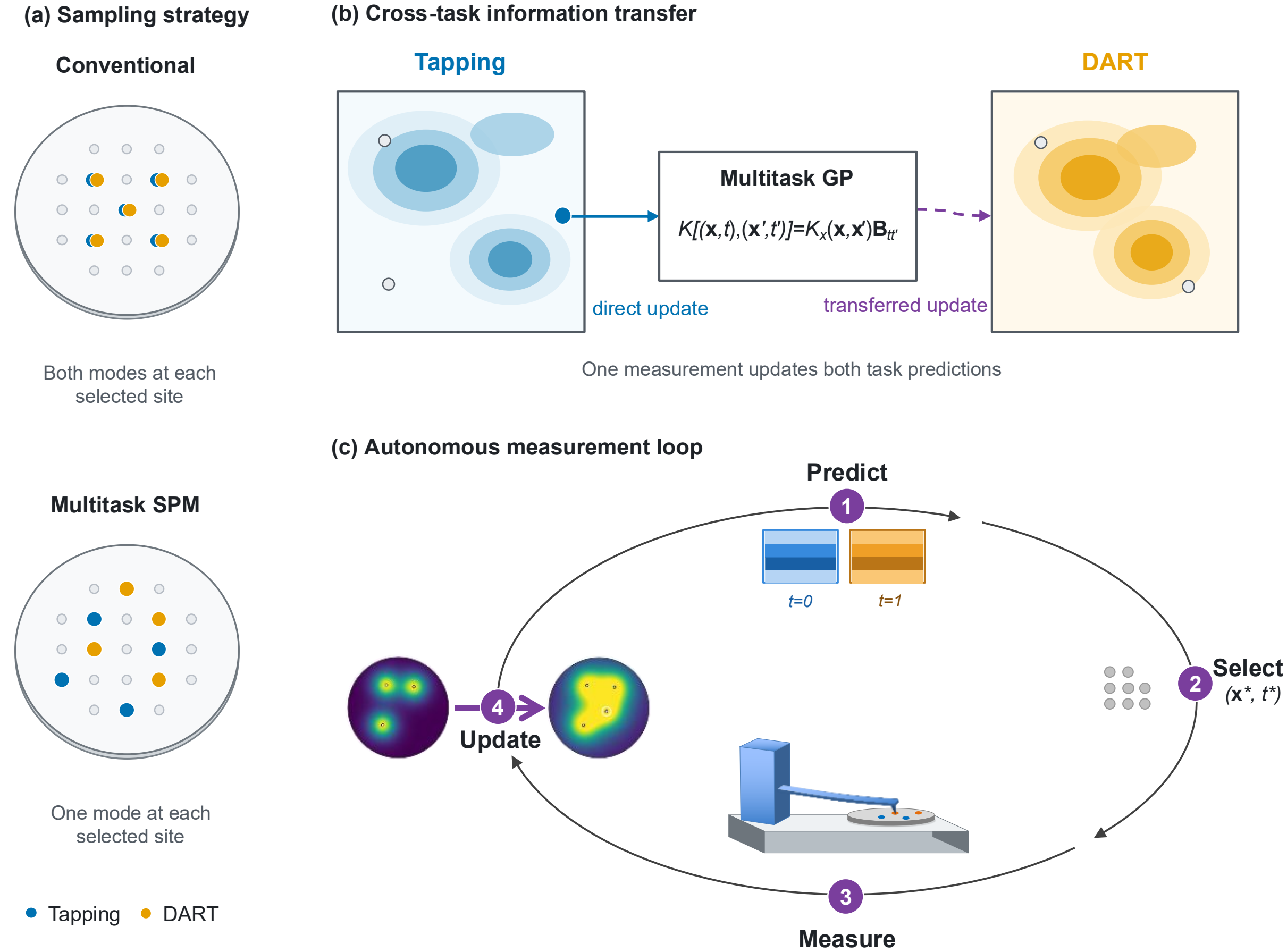

*Figure 1*. *Principle of multitask scanning probe microscopy. (a) Conventional SPM acquires both modes at every site; multitask SPM selects one mode per site, distributing the measurement budget across location and protocol. (b) A single tapping-mode measurement directly updates the tapping prediction and transfers information to the DART prediction through the learned task-covariance matrix* $\mathbf{B}$*. (c) Autonomous measurement loop: predict property maps for both tasks (1), select the next location-task pair (*$\mathbf{x}^*$*, $t^*$) (2), execute the measurement (3), and update the joint model (4).*

Several detector channels are commonly recorded in parallel during each protocol. A DART-PFM scan, for example, can simultaneously return topography, amplitude, phase, resonance frequency, quality factor, deflection, and error channels.[30-32] These channels form a multicomponent observation associated with one measurement task. Their simultaneous availability does not create a task-selection problem because acquisition of one channel does not exclude acquisition of the others. The multitask decision emerges when the microscope must choose among measurement protocols that cannot be performed simultaneously or that introduce distinct time or perturbation costs.

Let the spatial coordinate of the experiment be denoted by $\mathbf{x}$, and let $t$ denote a measurement task. The measured scalar response is written as

$$y_t(\mathbf{x}) = f_t(\mathbf{x}) + \varepsilon_t \quad (1)$$

where $f_t(\mathbf{x})$ is the unknown task-dependent response landscape and $\varepsilon_t$ represents measurement noise.[33] The response may correspond directly to a physical observable, such as current or surface potential, or to a descriptor extracted from a higher-dimensional measurement, such as image roughness, coercive voltage, loop area, magnetic contrast, or a feature derived from a local spectrum.

The purpose of the multitask model is to learn both the variation of $f_t$ over the experimental domain and the statistical relationship between different tasks. After the initial measurements, observations from the different tasks need not be obtained at the same positions. A tapping-mode measurement at one location and a contact-mode measurement at another location can be combined within the same posterior model.[26, 27]

**II.B. Intrinsic Coregionalization Model**

The present implementation uses the intrinsic coregionalization model (ICM).[26, 27] The augmented model input appends a task index to the spatial coordinate:

$$x_{\text{aug}} = [x_1, x_2, \dots, x_D, t] \quad (2)$$

The covariance between two location-task pairs is

$$K[(\mathbf{x}, t), (\mathbf{x}', t')] = K_x(\mathbf{x}, \mathbf{x}') \cdot \mathbf{B}_{tt'} \quad (3)$$

where $K_x$ is the spatial covariance kernel and $\mathbf{B}$ is the $T \times T$ task-covariance matrix. The task-covariance matrix is represented as

$$\mathbf{B} = \mathbf{W}\mathbf{W}^T + diag(\boldsymbol{\kappa}) \quad (4)$$

where $\mathbf{W}$ is a $T \times R$ matrix defining the low-rank shared latent representation and $\boldsymbol{\kappa} \in \mathbb{R}^T$ accounts for task-specific variance.[26] The rank $R$ controls the number of shared latent processes; in the present implementation $R = 2$. Task-dependent mean offsets and task-dependent noise terms are included to account for systematic differences between the measurement protocols. These offsets are treated as model parameters rather than removed by per-task standardization, and paired measurements are acquired at the seed positions, both of which improve identifiability of the cross-task correlation in multi-task Gaussian processes.[34]

The spatial kernel $K_x$ here is chosen to be a Matérn 5/2 kernel with automatic relevance determination,[33] providing independent length scales along the two wafer spatial coordinates:

$$K_x(\mathbf{x}, \mathbf{x}') = \sigma_f^2 \left(1 + \sqrt{5}\, r + \frac{5r^2}{3}\right) \exp(-\sqrt{5}\, r), \quad r = \sqrt{\sum_d \frac{(x_d - x'_d)^2}{\ell_d^2}} \tag{5}$$

where $\ell d$ is the length scale along spatial dimension $d$ and $\sigma^2_f$ is the output scale. The Matérn 5/2 kernel is preferred over the squared-exponential kernel[33, 35] because it permits less smooth spatial variation, which is appropriate for surface properties that may change sharply across the wafer.

This model assumes that the tasks share a common spatial correlation geometry.[26, 27] The individual response maps can have different amplitudes, offsets, and noise levels, while their spatial variations are governed by the same kernel length scales. This assumption is appropriate for the present demonstration, where tapping-mode and DART roughness are expected to reflect a common underlying surface morphology, but can be relaxed using the more general models described below.

## II.C. Alternative Multitask Models

The intrinsic coregionalization model represents one member of a broader family of multitask Gaussian processes.[27] At one limiting case, separate Gaussian processes can be assigned to each measurement protocol,

$$K_{tt'}(\mathbf{x}, \mathbf{x}') = 0, \quad t \neq t' \tag{6}$$

with task-specific kernels $K_t$. This formulation allows different spatial length scales and smoothness for each protocol but does not transfer information between tasks. In this case, the exploration campaigns for the individual tasks do not share information, and selection among the channels must instead be based on a common reward function (e.g., predictive uncertainty) combined with a suitable selection policy, such as $\varepsilon$-greedy as demonstrated for hypothesis learning.[20, 36, 37]

A more general linear model of coregionalization allows $Q$ different spatial kernels to contribute with task-specific weights[27]:

$$K_{tt'}(\mathbf{x}, \mathbf{x}') = \sum_{q=1}^{Q} \mathbf{B}_{tt'}^{(q)} \cdot K_q(\mathbf{x}, \mathbf{x}') \tag{7}$$

This construction can represent, for example, a long-range wafer gradient shared by all modalities and a shorter-range component visible primarily in a contact or electrical measurement.

An alternative physically interpretable representation separates the response into shared and task-specific components,[27, 38]

$$f_t(\mathbf{x}) = a_t f_{shared}(\mathbf{x}) + \delta_t(\mathbf{x}) \quad (8)$$

where $f_{\text{shared}}$ represents material structure expressed across all modalities and $\delta_t$ represents modality-specific physical information, contact effects, or measurement artifacts. These more general models introduce additional parameters and may require more observations for stable fitting. Correspondingly, the intrinsic coregionalization model provides a suitable first implementation for real-time experimental control, while the linear-coregionalization and shared-plus-task-specific models provide a route toward analysis of more weakly correlated AFM modalities.

Note that the selection of the proper algorithm as well as selection of the non-zero mean function[21, 39, 40] for the Gaussian Process is deeply domain specific and depends on a particular system, available physical knowledge, and imaging task.

**II.D. Selection of Location and Measurement Task**

At each active-learning step, the model predicts the posterior mean and uncertainty for every task at all candidate locations (Fig. 1c). The decision domain is the joint set of unmeasured location-task pairs[41]:

$$\mathcal{D} = \{ (\mathbf{x}_i, t_j) \} \quad (9)$$

The present implementation uses a random-scalarization upper-confidence-bound acquisition function, inspired by ParEGO but using a randomly weighted sum in place of the augmented Tchebycheff scalarization.[42, 43] At each step, task weights $\mathbf{w}$ are sampled from a symmetric Dirichlet distribution,[43]

$$\mathbf{w} \sim \text{Dirichlet}(\mathbf{1}_T) \quad (10)$$

and applied to a standardized upper-confidence-bound (UCB) value[44] computed separately for each task. For tasks where a lower response is preferred, the sign of the predicted mean is reversed before scoring. The selected experiment is

$$(\mathbf{x}^*, t^*) = \arg \max_{(\mathbf{x},t)\in\mathcal{D}} w_t \left( \text{sign}_t \cdot \hat{\mu}_t(\mathbf{x}) + \sqrt{\beta_{\text{eff}}} \cdot \sigma_t(\mathbf{x}) \right) \quad (11)$$

where $\hat{\mu}_t(\mathbf{x})$ and $\sigma_t(\mathbf{x})$ are the standardized posterior mean and standard deviation for task $t$ at location $\mathbf{x}$, $sign_t = -1$ for minimization tasks and +1 for maximization tasks, and $\beta_{\text{eff}}$ is an exploration parameter that is annealed over the course of the experiment:

$$\beta_{\text{eff}} = \frac{\beta_0}{1 + 0.15 \cdot n_{\text{step}}}, \quad \beta_0 = 2.0 \tag{12}$$

with $n_{\text{step}}$ is the current iteration count. Annealing progressively shifts the acquisition toward exploitation as the model becomes better informed. The random Dirichlet draw couples task selection to spatial selection: each step's weight vector determines which task's uncertainty contributes most to the acquisition score, producing a stochastic task-allocation policy without requiring an explicit cost model.

To prevent spatial clustering and biased hyperparameter estimation, a random exploration step is inserted every three active-learning iterations. At these steps, the framework selects the candidate location farthest from all previously measured positions,[45] with task assigned uniformly at random. The GP hyperparameters are still re-optimized at exploration steps, so the model remains current throughout.

For measurement protocols with strongly unequal acquisition times or perturbation levels, the acquisition function can be extended to include explicit cost,[35, 41, 46]

$$a_{\text{cost}}(\mathbf{x}, t) = \frac{\Delta I(\mathbf{x}, t)}{c_{\text{time},t} + \lambda\, c_{\text{damage},t} + \gamma\, c_{\text{switch},t}} \tag{13}$$

where $\Delta I$ is the expected information gain and the denominator represents task-dependent acquisition time, sample or tip perturbation, and mode-switching overhead. Such cost-aware task allocation is particularly relevant for combinations of rapid imaging and local spectroscopy. The present implementation uses equal implicit cost for both tasks; Equation 13 is described here as a natural extension.

## III. Experimental Implementation

### III.A. Automated Scanning Probe Platform

Multitask scanning probe microscopy was implemented on an Oxford Instruments Asylum Research Jupiter AFM equipped with a motorized large-sample stage and software-accessible instrument controls. The microscope was interfaced through the aespm automated-experiment software environment,[23] enabling programmatic control of stage motion, microscope-mode selection, probe tuning, surface approach, scan execution, data transfer, and scalar descriptor extraction.

The large spatial range of the wafer introduces several practical requirements absent in conventional local active-learning experiments. First, stage translation changes the local sample height and requires compensation for wafer tilt and bow. A reference height map was constructed from height measurements at 17 locations distributed across the wafer. A radial basis function interpolant (thin-plate spline)[47] was fitted to these reference points and used to provide the initial sample-height estimate at each new candidate position before approach.

Switching between tapping and DART modes requires loading the corresponding instrument profile, tuning the cantilever to the appropriate resonance — single-frequency AC tune for tapping, dual-frequency DART tune for contact resonance,[30, 32] setting the appropriate setpoint and feedback gain conditions, and executing a fresh surface approach. All of these steps were automated within the closed-loop workflow.

### III.B. Model System and Experimental Domain

An AlScN composition-spread wafer was selected as the model system.[28, 29] The experimental domain was represented as a circular region with a radius of 35 mm. A hexagonal grid with 2 mm node spacing generated 1,117 candidate measurement positions across the wafer (Fig. 2a). Grid coordinates were normalized to the unit square for Gaussian-process training, while all stage movements and graphical representations retained physical coordinates in millimeters.

The present experiment defines two tasks:

$$t = 0: \quad \text{tapping} - \text{mode roughness} \tag{14}$$

$$t = 1: \quad \text{DART} - \text{mode roughness} \tag{15}$$

For each acquired image, the height channel was converted to nanometers. The scalar roughness descriptor was computed as the standard deviation of the height values after excluding pixels more than five standard deviations from the image mean. This filtering suppresses the effect of isolated contamination particles, feedback transients, and other outlier pixel values on the scalar reward. The two tasks provide protocol-dependent measurements of the same nominal surface property. Tapping-mode imaging probes the surface through intermittent contact[12] whereas the DART measurement is acquired under sustained contact with contact-resonance feedback.[30, 31]

### III.C. Model Training and Hyperparameter Optimization

All ICM model parameters — spatial kernel length scales $\{\ell_d\}$, output scale $\sigma_f^2$, task covariance factors $\{\mathbf{W}, \boldsymbol{\kappa}\}$, per-task noise $\{\sigma_t^2\}$, and per-task mean offsets $\{\mu_t\}$ — were optimized jointly by maximizing the exact log marginal likelihood[33] using the Adam optimizer.[48] At the initialization stage and at each full retraining step, 300 optimization iterations were run with a learning rate of 0.01. A warm-start scheme was used for subsequent iterations: if a previously trained model exists, the optimizer is run for 50 iterations starting from the previous parameter values.

All GP computations used the GPyTorch library[49] with exact Cholesky-based inference, tractable at the observation counts encountered here ($N_{\text{obs}} \lesssim 100$; 35 in the present experiment). Predictions were evaluated in batches of 5,000 candidate locations to manage GPU memory. Posterior means and standard deviations reported for the final landscapes were evaluated from the converged model hyperparameters applied to the complete set of 35 observations. The standard deviations quoted are those of the latent response, excluding the additive measurement-noise term, so that they represent uncertainty in the reconstructed surface rather than in a prospective new measurement.

### III.D. Initialization and Closed-Loop Experiment

Five seed positions were selected uniformly at random from the candidate grid (random seed 247; Fig. 2b). Both tapping and DART measurements were performed at each seed location, providing 10 initial location-task observations at 5 per task. These paired measurements establish an initial estimate of the task-covariance matrix **B** before decoupled active selection begins.
Following initialization, only one task was measured at each selected location. From this point onward (the randomly drawn seed positions being exempt), the spatial edge of the wafer was penalized using a soft Gaussian mask that suppresses scores within approximately 10 mm of the wafer rim, with a hard exclusion within 3 mm. Every third active-learning step was designated as a forced exploration step (Fig. 2c). The experiment ran for 25 active-learning steps following the 5 seed locations, for a total of 35 individual scans.

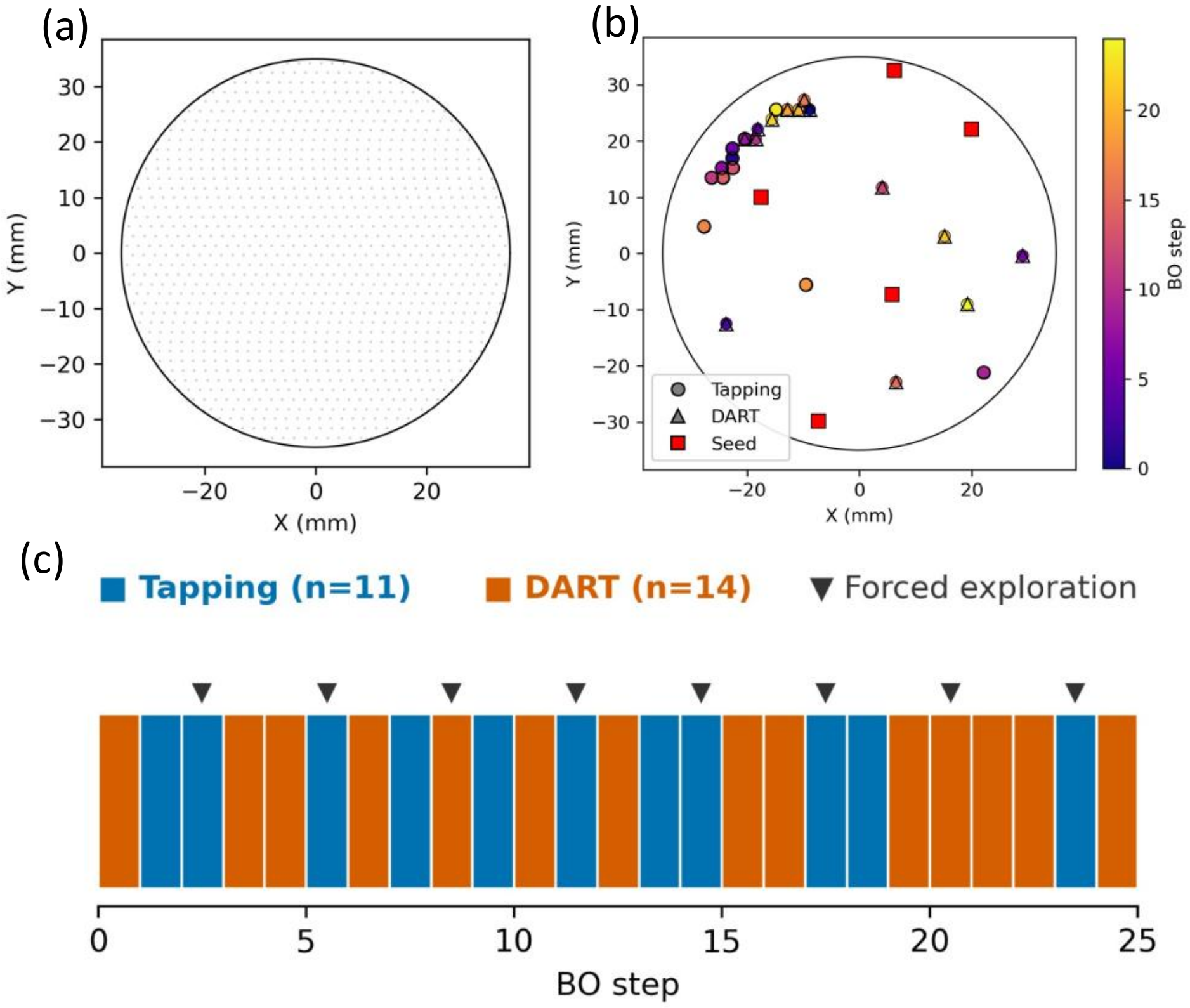

***Figure 2***. *Experimental implementation and model system. (a) Hexagonal candidate grid of 1,117 positions spanning a 35 mm radius wafer at 2 mm node spacing. (b) Measurement trajectory over the 35 individual scans that make up the experiment: 10 seed scans (5 positions measured in both modes, red squares) followed by 25 active-learning scans (one mode each). Points are colored by iteration order; circles indicate tapping-mode and triangles indicate DART measurements. (c) Task selected at each of the 25 active-learning steps. The 8 forced-exploration steps (every third iteration), at which the task is assigned uniformly at random, are marked by triangles. Over the 25 active steps DART was selected somewhat more often (n=14) than tapping (n=11); the difference is within the range expected by chance for this number of steps.*

### III.E. Learning the Relationship Between Tapping and DART Measurements

The initial paired measurements establish the relationship between roughness obtained in tapping and DART modes. Both responses vary across the wafer, reflecting the spatially nonuniform surface morphology (Figs. 3d,e). The DART measurements exhibit a systematic difference in magnitude relative to tapping-mode measurements, consistent with the different probe-surface interaction and feedback conditions[31, 32] (Fig. 3b). The learned task-correlation matrix yields $\rho = B12/\sqrt{B_{11}B_{22}} = 0.752$, indicating substantial positive correlation between the two modes (Fig. 3a). This is a property of the fitted model rather than of the raw data: the Pearson coefficient of the measured-versus-predicted cross-task scatter is r = 0.304 (Fig. 3b), the difference reflecting the per-measurement noise (Fig. 4e) and the small number of paired observations. The five co-located seed pairs are the primary constraint on ρ; co-located measurements are known to improve identifiability of the cross-task correlation, which is otherwise recovered only slowly from non-overlapping designs.[34] The resulting posterior mean maps are strongly co-varying across the candidate grid (r = 0.984, Fig. 3c); because both maps derive from one joint model with a shared spatial kernel, this agreement is partly imposed by construction and should not be read as independent validation. The spatial difference between the two predictions is shown in Fig. 3f.

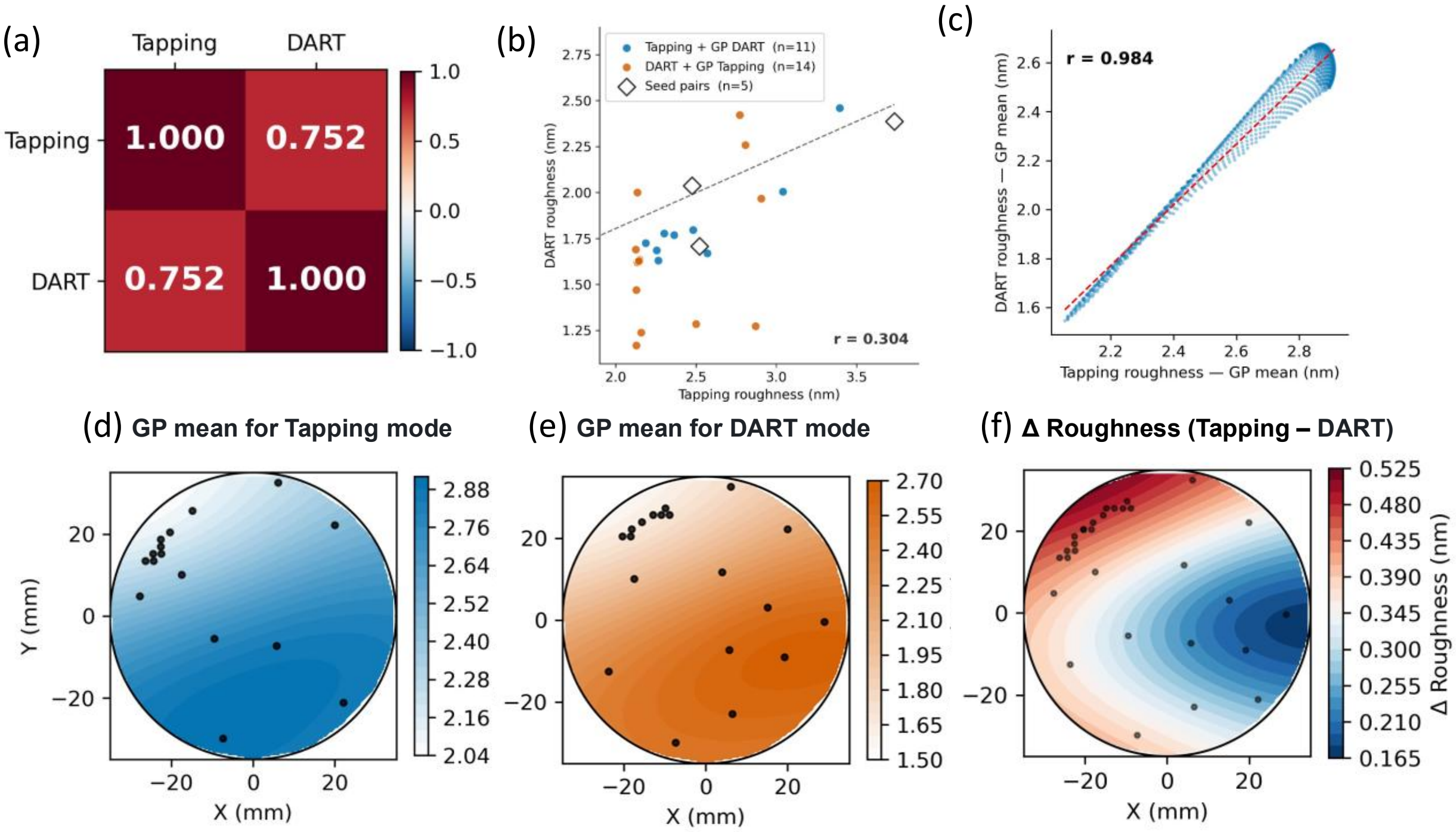

***Figure 3.*** *Learned relationship between tapping and DART tasks. (a) Learned task-correlation matrix **B**, with off-diagonal entry ρ = B12/(B11B22)1/2 = 0.752, the learned correlation between the two measurement modes. (b) Cross-task scatter of measured roughness values; blue points*

*show tapping measurements paired with GP-predicted DART roughness at the same location, orange points show DART measurements paired with GP-predicted tapping roughness, and diamonds mark the five paired seed measurements. The Pearson coefficient of this measured-versus-predicted scatter ($r$ = 0.304) is much lower than the learned task correlation in (a), reflecting the per-measurement noise and the small number of samples rather than an absence of cross-task structure. (c) Scatter plot of GP posterior mean roughness at all 1,117 candidate grid locations ($r$ = 0.984); because both maps are generated by a single joint model with a shared spatial kernel, this agreement is largely imposed by the model and is not an independent validation. (d) GP posterior mean roughness map for tapping mode across the wafer; black dots mark the 16 locations measured in tapping mode. (e) GP posterior mean roughness map for DART mode; black dots mark the 19 locations measured in DART mode. (f) Difference between the tapping and DART posterior means. The offset is positive across the whole wafer and varies along the upper-left to lower-right diagonal, from approximately 0.53 nm in the upper left to approximately 0.17 nm toward the lower right.*

### III.F. Autonomous Selection of Measurement Modality

Following the paired seed measurements, the controller selected individual location-task pairs according to the random-scalarization UCB acquisition function (Fig. 4a,b). Each task-specific measurement updates the joint posterior. A DART measurement changes the predicted DART map directly and changes the tapping prediction according to the learned task covariance. This information transfer is the defining feature of multitask sampling:[26] the experiment can construct two response landscapes without acquiring both measurement modes at every position. Over the 25 active steps the controller selected DART 14 times and tapping 11 times (Fig. 4a,b); this difference is within the range expected by chance, and 8 of the 25 assignments were made by the forced-exploration rule rather than by the model. The learned per-task noise differs substantially (0.233 nm for tapping and 0.437 nm for DART, Fig. 4e), and the running standard deviation of the observations for each task is shown in Fig. 4c. Averaged over the candidate grid, the posterior standard deviation after all 35 measurements is 0.176 nm for tapping and 0.235 nm for DART, against prior standard deviations of 0.925 and 1.098 nm respectively (Fig. 4f); the joint model therefore removes roughly 80% of the prior uncertainty in both landscapes while measuring only one modality at each location.

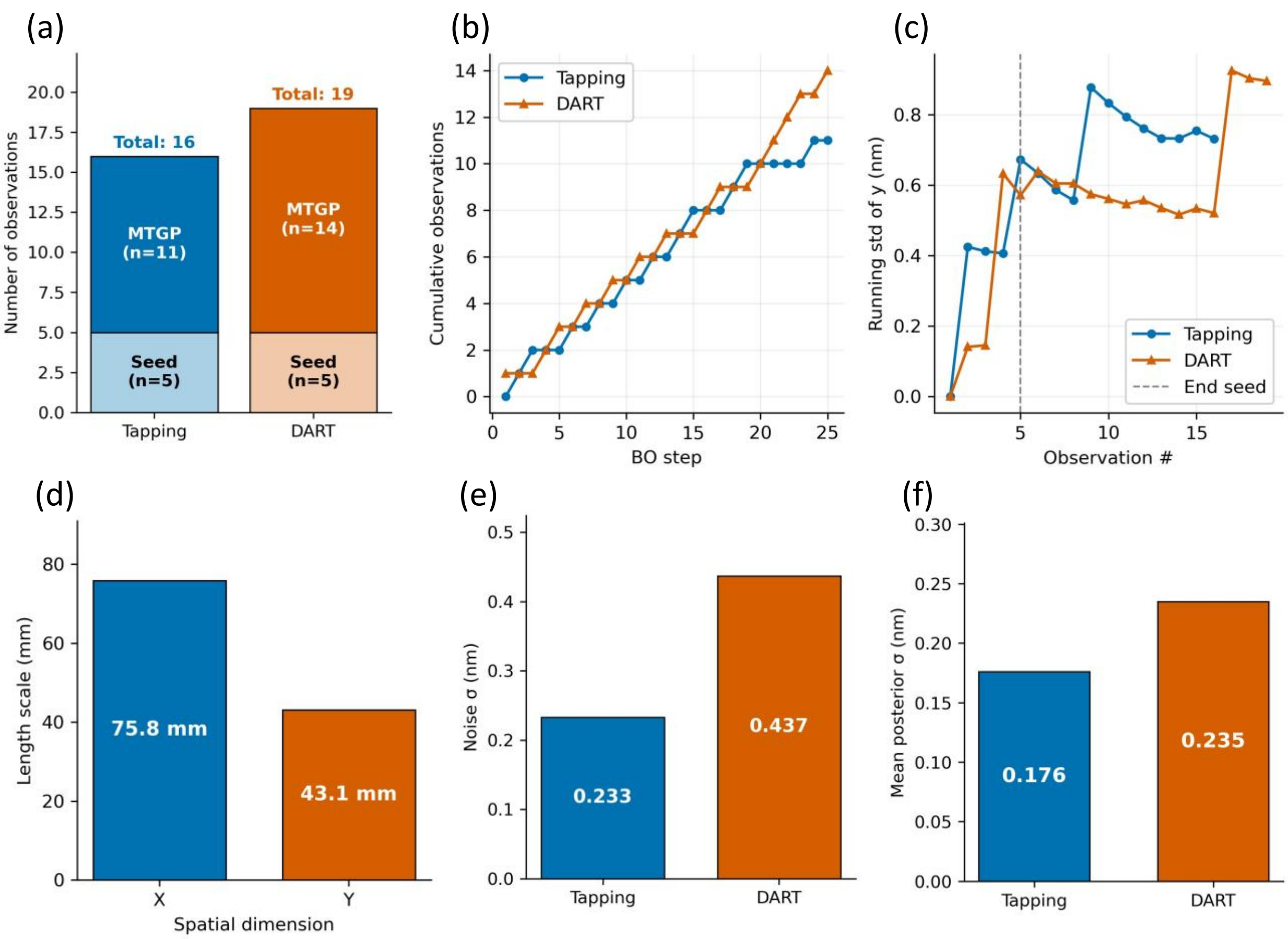

***Figure 4.*** *Autonomous task allocation and reconstruction performance. (a) Total observations per task (16 tapping, 19 DART), split into the 5 paired seed observations and the 25 active-learning observations; of the latter, 17 were selected by the multitask Gaussian process and 8 were assigned by the forced-exploration rule. (b) Cumulative task observations over the 25 active-learning steps, excluding the seed phase. (c) Running standard deviation of the observed roughness values for each task; the dashed line marks the end of the seed phase at five observations per task. (d) Learned spatial length scales along the X and Y wafer dimensions. (e) Learned per-task measurement noise: DART is approximately twice as noisy as tapping (0.437 nm versus 0.233 nm). (f) Mean posterior standard deviation across the candidate grid after all 35 measurements.*

## IV. From Correlated Topographic Measurements to Functional Multitask SPM

The tapping-DART experiment provides a controlled implementation of multitask scanning probe microscopy. Both tasks are expected to share a strong underlying spatial structure, allowing the instrumentation, model training, mode selection, and joint prediction to be tested in a favorable transfer-learning regime. The same framework becomes particularly consequential when the available protocols differ substantially in acquisition cost and physical information content. One natural extension combines tapping-mode topography with contact current mapping.[2, 25] Tapping measurements can be distributed broadly across a wafer or combinatorial library, while contact current measurements are deployed in regions where morphology does not provide a sufficiently certain prediction of electrical behavior.

A second implementation combines fixed-bias current mapping with local current-voltage spectroscopy.[2] The multitask controller can use dense fixed-bias measurements and sparse voltage sweeps to reconstruct the nonlinear transport landscape. For ferroelectric materials, small-signal PFM imaging and local switching spectroscopy form a natural task hierarchy.[4, 13] A multitask model can determine where the equilibrium electromechanical response predicts switching and where direct switching measurements remain necessary.

KPFM and conductive AFM provide related equilibrium and nonequilibrium electrical observables.[2, 3] Their joint analysis can identify locations where surface potential and current transport are correlated and regions where electrostatic and transport behavior decouple.

Explicit representation of measurement cost is equally important. The experimental value of multitask SPM emerges when an inexpensive, rapid, or weakly perturbative measurement can reduce the number of slower contact or spectroscopic measurements. Incorporating these quantities into the acquisition function via Eq. 13 will allow the microscope to select measurements according to their expected scientific value per unit experimental cost.[35, 41, 46]

## V. Conclusions

We have introduced and experimentally operationalized multitask scanning probe microscopy in a live closed loop, extending autonomous SPM from selection of spatial coordinates to autonomous selection and execution of combined location-measurement pairs. The framework uses a multitask Gaussian process with an intrinsic coregionalization model to learn the spatial response of individual AFM protocols together with their cross-task covariance. Measurements acquired using one protocol can therefore update predictions for the other protocols, allowing the instrument to distribute its measurement budget across both spatial and modality dimensions.

The approach was implemented on an automated large-sample AFM and demonstrated for tapping and DART roughness measurements across a composition-spread AlScN wafer. Five paired seed positions established the initial task relationship (Fig. 3a,b), after which the microscope executed 25 further individual task-location combinations, 17 of which were selected autonomously by the multitask model and 8 assigned by the periodic forced-exploration rule (Fig. 4a,b). The intrinsic coregionalization model captured the common spatial structure of the two measurements (Figs. 3c–e) and enabled joint reconstruction from noncoincident observations.

The present experiment establishes the instrumental and computational basis for multitask SPM. Extension to combinations of topography, current mapping, local current-voltage spectroscopy, KPFM, PFM, switching spectroscopy, and magnetic-force measurements will enable rapid and weakly perturbative modes to guide selective deployment of slower, contact-based, or state-changing measurements.

**Acknowledgements**

The work of A.R., Y.L., and S.V.K. on the development and implementation of the multitask Gaussian-process framework and autonomous scanning probe microscopy workflow was supported by the U.S. National Science Foundation, Division of Materials Research, Ceramics Program, under Award No. 2523284. The work of JP.M. and I.M. on the fabrication of the AlScN sample was supported by the Center for 3-Dimensional Ferroelectric Microelectronics Manufacturing (3DFeM2) under Award No. DE-SC0021118.

**Data Availability**

The code implementing the multitask Gaussian process framework and the closed-loop measurement workflow is openly available at https://github.com/adityaraghavan98/multitask-spm.